\documentclass[sigconf]{acmart}

\AtBeginDocument{%
  \providecommand\BibTeX{{%
    \normalfont B\kern-0.5em{\scshape i\kern-0.25em b}\kern-0.8em\TeX}}}

\copyrightyear{2023}
\acmYear{2023}
\setcopyright{acmlicensed}\acmConference[WWW '23 Companion]{Companion Proceedings of the ACM Web Conference 2023}{April 30-May 4, 2023}{Austin, TX, USA}
\acmBooktitle{Companion Proceedings of the ACM Web Conference 2023 (WWW '23 Companion), April 30-May 4, 2023, Austin, TX, USA}
\acmPrice{15.00}
\acmDOI{10.1145/3543873.3584656}
\acmISBN{978-1-4503-9419-2/23/04}

\usepackage[inline]{enumitem}
\usepackage{caption}
\usepackage{subcaption}
\begin{document}

\title{PIE: Personalized Interest Exploration for Large-Scale Recommender Systems}


\author{Khushhall Chandra Mahajan}
\authornote{Authors contributed equally to this research.}
\affiliation{%
  \institution{Meta Inc., USA}
  \streetaddress{}
  \city{}
  \state{}
  \country{}
  \postcode{}
}
\email{khushhall@meta.com}

\author{Amey Porobo Dharwadker}
\authornotemark[1]
\affiliation{%
  \institution{Meta Inc., USA}
  \streetaddress{}
  \city{}
  \state{}
  \country{}
  \postcode{}
}
\email{ameydhar@meta.com}

\author{Romil Shah}
\authornotemark[1]
\affiliation{%
  \institution{Meta Inc., USA}
  \streetaddress{}
  \city{}
  \state{}
  \country{}
  \postcode{}
}
\email{romilshah@meta.com}

\author{Simeng Qu}
\authornotemark[1]
\affiliation{%
  \institution{Meta Inc., USA}
  \streetaddress{}
  \city{}
  \state{}
  \country{}
  \postcode{}
}
\email{simonequ@meta.com}

\author{Gaurav Bang}
\authornotemark[1]
\affiliation{%
  \institution{Meta Inc., USA}
  \streetaddress{}
  \city{}
  \state{}
  \country{}
  \postcode{}
}
\email{gauravbang@meta.com}
\author{Brad Schumitsch}
\affiliation{%
  \institution{Meta Inc., USA}
  \streetaddress{}
  \city{}
  \state{}
  \country{}
  \postcode{}
}
\email{bschumitsch@meta.com}

\renewcommand{\shortauthors}{Mahajan, et al.}

\begin{abstract}
Recommender systems are increasingly successful in recommending personalized content to users. However, these systems often capitalize on popular content. There is also a continuous evolution of user interests that need to be captured, but there is no direct way to systematically explore users' interests. This also tends to affect the overall quality of the recommendation pipeline as training data is generated from the candidates presented to the user. In this paper, we present a framework for exploration in large-scale recommender systems to address these challenges. It consists of three parts, first the user-creator exploration which focuses on identifying the best creators that users are interested in, second the online exploration framework and third a feed composition mechanism that balances explore and exploit to ensure optimal prevalence of exploratory videos. Our methodology can be easily integrated into an existing large-scale recommender system with minimal modifications. We also analyze the value of exploration by defining relevant metrics around user-creator connections and understanding how this helps the overall recommendation pipeline with strong online gains in creator and ecosystem value. In contrast to the regression on user engagement metrics generally seen while exploring, our method is able to achieve significant improvements of 3.50\% in strong creator connections and 0.85\% increase in novel creator connections. Moreover, our work has been deployed in production on Facebook Watch, a popular video discovery and sharing platform serving billions of users.
\end{abstract}



\maketitle

\section{Introduction}

Large-scale recommender systems run the risk of creating a feedback loop where interaction labels in the training data are generated primarily based on items previously recommended to users. Hence it is important that exploration strategies are built into the system to maintain the value of the system to users. Exploration in large-scale recommendation systems helps users continually find relevant content by better understanding their interests.

In this paper, we present a framework for introducing exploration in our real-world large-scale recommendation system on Facebook Watch. This framework easily integrates with an existing deployed system. The paper is divided into three parts: 

 \begin{itemize}
\item Personalized Exploration Space: We actively try to find creators with high affinity towards our users. We use the Personalized PageRank (PPR) algorithm, a random walk based, non-parametric method for finding a set of neighboring nodes for each node in a graph to discover new creators. More details are explained in Section \ref{Personalized-Exploration-Space}.
\item Online Exploration Framework: Traversing an uncertain interest space to identify new interests often leads to decrease in users relevance. We formulate this problem as an online bandit problem and in Section \ref{Online-Exploration} discuss implementations that help us achieve efficient online exploration with minimal user engagement loss.
\item Delivery to Balance Explore/Exploit: Integrating exploratory content into a user's feed ordered by point-wise ranker requires special ranking stack modules.  We discuss in Section \ref{Delivery-Balance} a method used to combine relevance and exploration value for optimal long-term rewards.
\end{itemize}

Facebook Watch is one of the largest global destinations for discovering and sharing video contents, with more than 1.25 billion monthly viewers \cite{FBWatchDap}. Videos are recommended to users by our personalized discovery engine, which continually tries to serve the best content to users based on their interests, new trends, creators affinity, etc. Our design is similar to a two-stage recommender system \cite{covington2016deep} and has two main parts, retrieving relevant candidate items and ranking the retrieved items in descending relevance order.

\section{Related Work}
In recent years, recommendation systems that provide users with personalized content by extracting user preferences from observed interaction data have been very successful. However, such systems tend to reinforce or support users’ current preferences \cite{ekstrand2016behaviorism} and tend to reinforce popularity bias \cite{abdollahpouri2019popularity}. For example, a user might watch a recommended video just because it’s popular and many other users have watched it, e.g. On-demand video recommendation platforms often display videos with their number of views and it could influence user interactions. In fact, these interactions are primarily driven by the popularity of the recommended videos rather than their relevance to the individual user.

Recently, also \cite{fabbri2022exposure} and \cite{cinus2022effect} have focused on feedback loops and long term effects of people recommender systems. The former analyzes inequalities in the exposure of minority groups according to their homophily level and the latter focuses on polarization and echo chambers. Exploration in recommender systems is essential to fulfill users’ needs for variety and novelty \cite{liang2019recommender} and address potential filter bubbles \cite{nagulendra2014understanding}, echo chambers \cite{baumann2020modeling}, and polarization \cite{dandekar2013biased} issues.

\section{Problem}
Recommender system models are often reliant on data in order to accurately solve problems. However, without data, these systems can be limited in their ability to find new interests or escape from local minima. For example, if a particular user is interested in basketball videos, and has never been exposed to that content, the machine learning system will have no training data to learn this from. In other words, we may not be aware of all the different interests that users could have. Additionally, user interests can change over time in response to recent trends or news events. Therefore, it is important for recommender systems to systematically explore a variety of content in order to serve relevant recommendations to users.

\begin{figure}[h]
  \centering
  \captionsetup{font=small}
  \addtolength{\abovecaptionskip}{-5pt}
  \addtolength{\belowcaptionskip}{-5pt}
  \includegraphics[width=0.75\linewidth]{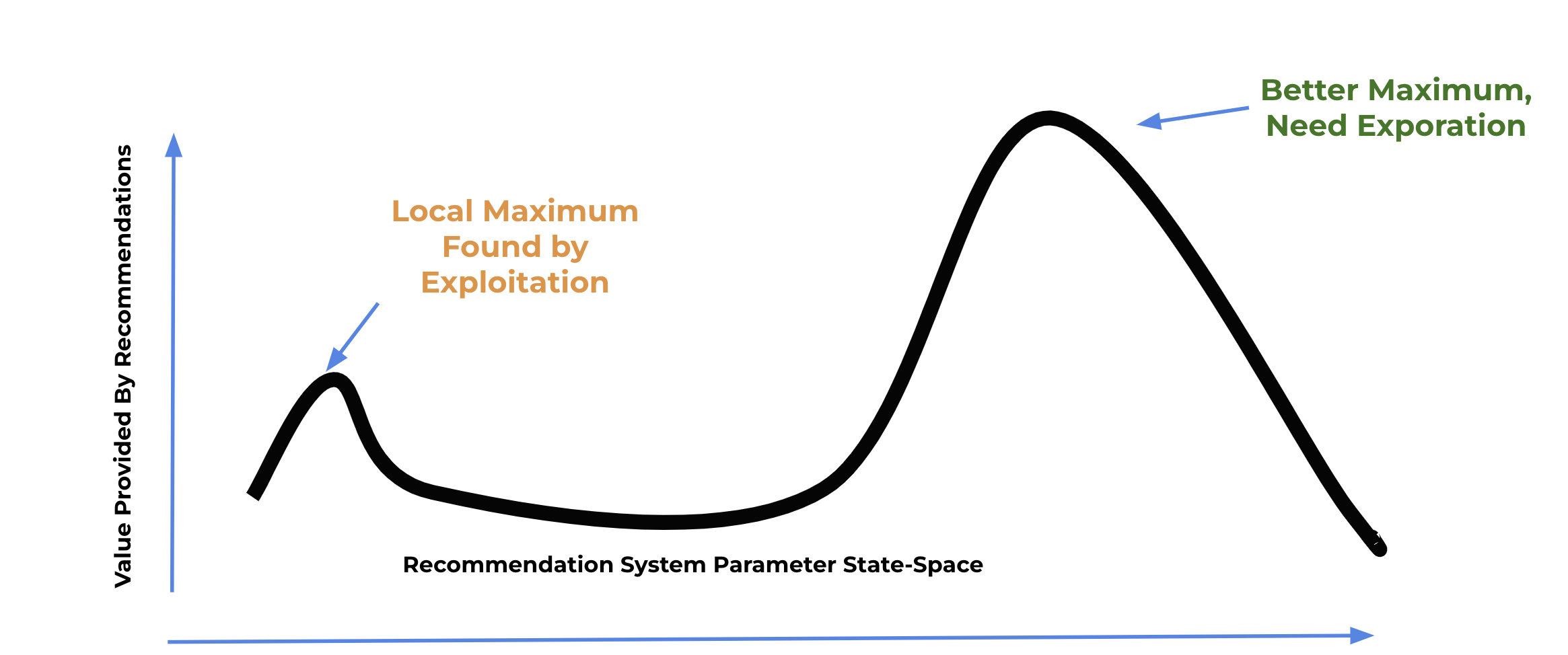}
  \caption{Exploration of users' interests could lead to a better maxima on the value provided by recommender systems.}
  \Description{A graph showing that exploitation of user interests leads to a local maxima on the value provided by recommender systems, while a global maxima can be attained through exploration.}
\end{figure}

Merely exploiting our user understanding leads to repetitive content that misses out on various other user interests. We want to explore other topics we're less certain about so we can find more interests, or introduce new ones to users. This exploration needs to be efficient, effective, and optimally balanced with our exploitation strategy.

\section{System Design}

\begin{figure}[h]
  \centering
  \captionsetup{font=small}
  \addtolength{\abovecaptionskip}{-5pt}
  \addtolength{\belowcaptionskip}{-5pt}
  \includegraphics[width=0.75\linewidth]{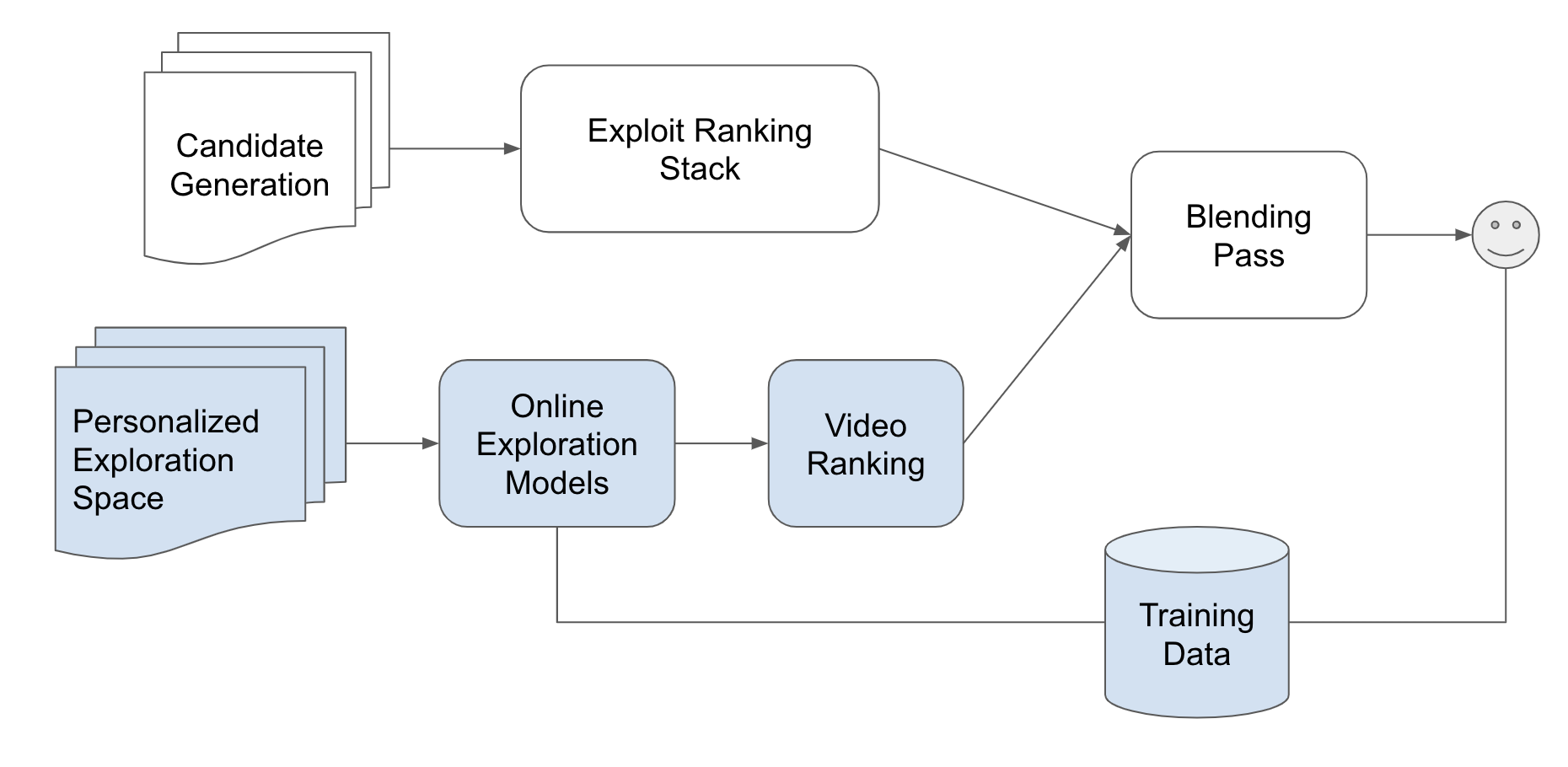}
  \caption{Incorporating personalized exploration in existing recommender system ranking framework.}\label{sysfig}
  \Description{Flow diagram showing how to incorporate personalized exploration in existing recommender system ranking framework.}
\vspace{-3mm}
\end{figure}

We design the exploration system on top of an existing recommendation system. The blue path in Figure \ref{sysfig} demonstrates all the major components. The exploration system interacts with the existing recommender system only at the blending pass, namely the best exploration videos after ranking are blended with non-exploration videos before serving users. The same design can easily be applied to any existing recommender system.

\subsection{Personalized Exploration Space} \label{Personalized-Exploration-Space}

Exploring the affinity between each user and each piece of content can be very exhaustive. Instead, we try to learn the affinity between each user and each attribute, which is a much smaller set. We currently focus on learning user-creator connections. 

\begin{figure}[h]
  \centering
  \captionsetup{font=small}
  \addtolength{\abovecaptionskip}{-5pt}
  \addtolength{\belowcaptionskip}{-5pt}
  \includegraphics[width=0.75\linewidth]{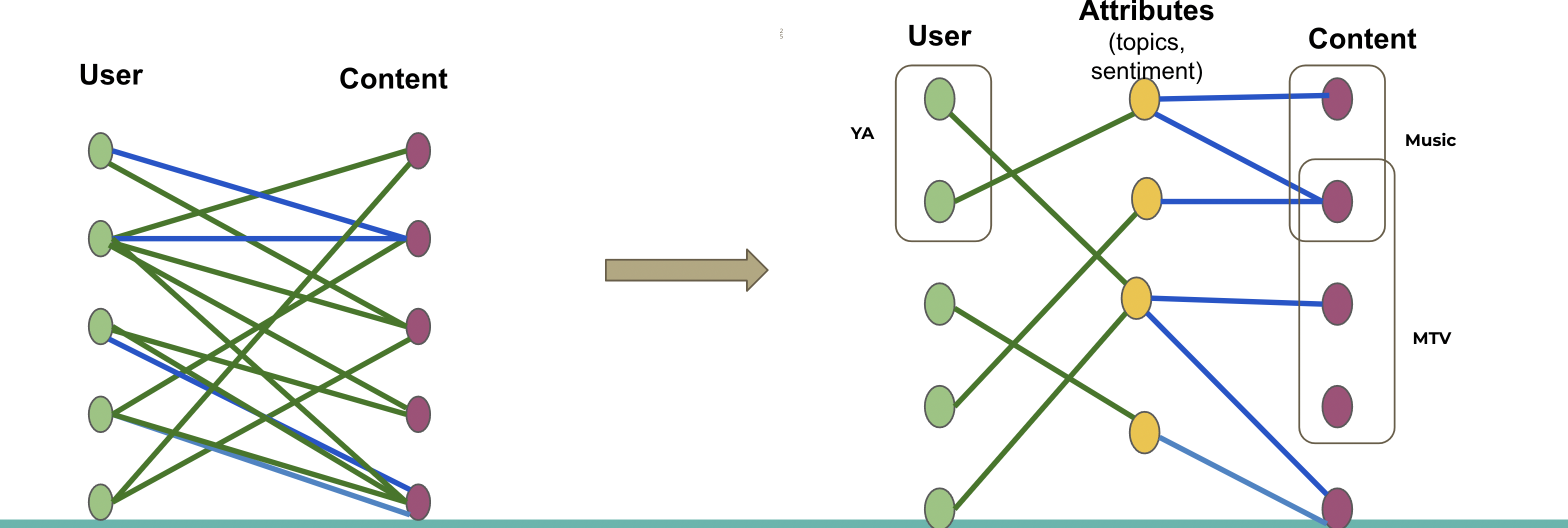}
  \caption{Moving from user-content to user-attribute-content}
  \Description{}
\end{figure}

We created user-creator affinity using the Personalized PageRank (PPR) algorithm. PPR is a random walk based graph learning algorithm that learns entity-entity similarities based on engagement data. Its advantage compared to collaborative filtering based algorithms is that PPR can leverage multi-hop information and therefore allows more exploration. Additionally, PPR is friendly to creators with low prior engagements.

Our approach is described as follows. First, we create a bi-graph between users and creators as source and target and vice versa. We then assign weights to edges based on user engagement over the past few days. We then run a personalized creator rank algorithm and filter out the nodes that are creators. By this step we obtain a set of similar creators for a given creator. Now, we take user interactions with creators and use the output from the last step to aggregate top-k creators. These creators are new to the user by design and fit our exploration use case very well. 

To improve exploration efficiency, we add two filters at the creator level. \textbf{Novelty Filter} removes creators that a user has previously interacted with, ensuring that the exploration system only captures user's new interests. \textbf{Quality Filter} removes low quality creators, such as creators with Watchbait content or creators that received low engagement at a aggregated level through exploration.

\subsection{Online Exploration} \label{Online-Exploration}
The goal of user exploration is to reduce uncertainty by recommending novel creators to users. This is done by starting with a high cardinality of options and converging towards a state where a novel creator is considered `expired' or `connected'. While there are opportunity costs associated with this process, users could potentially benefit from more diversified recommendations that adapt to their taste.


The process of exploring the space created for a user could be formulated as an online bandit problem. A simple solution could be Thompson sampling while a more refined solution could be contextual bandit where user, creator and interaction features are considered. To keep it simple, we stick with Thompson sampling here. Each time an exploratory video is presented, the goal is to maximize the cumulative reward based on past context and actions. We use Thompson sampling to find creators and then select videos from the selected creator. We define reward as the number of videos engaged from the creator and rest as failures. Rewards and rests are used to compute the parameters $\alpha$ and $\beta$ in Thompson sampling. This exploration happens at the user level.

\begin{figure}[h]
  \centering
  \captionsetup{font=small}
  \addtolength{\abovecaptionskip}{-2pt}
  \addtolength{\belowcaptionskip}{-2pt}
  \includegraphics[width=0.9\linewidth]{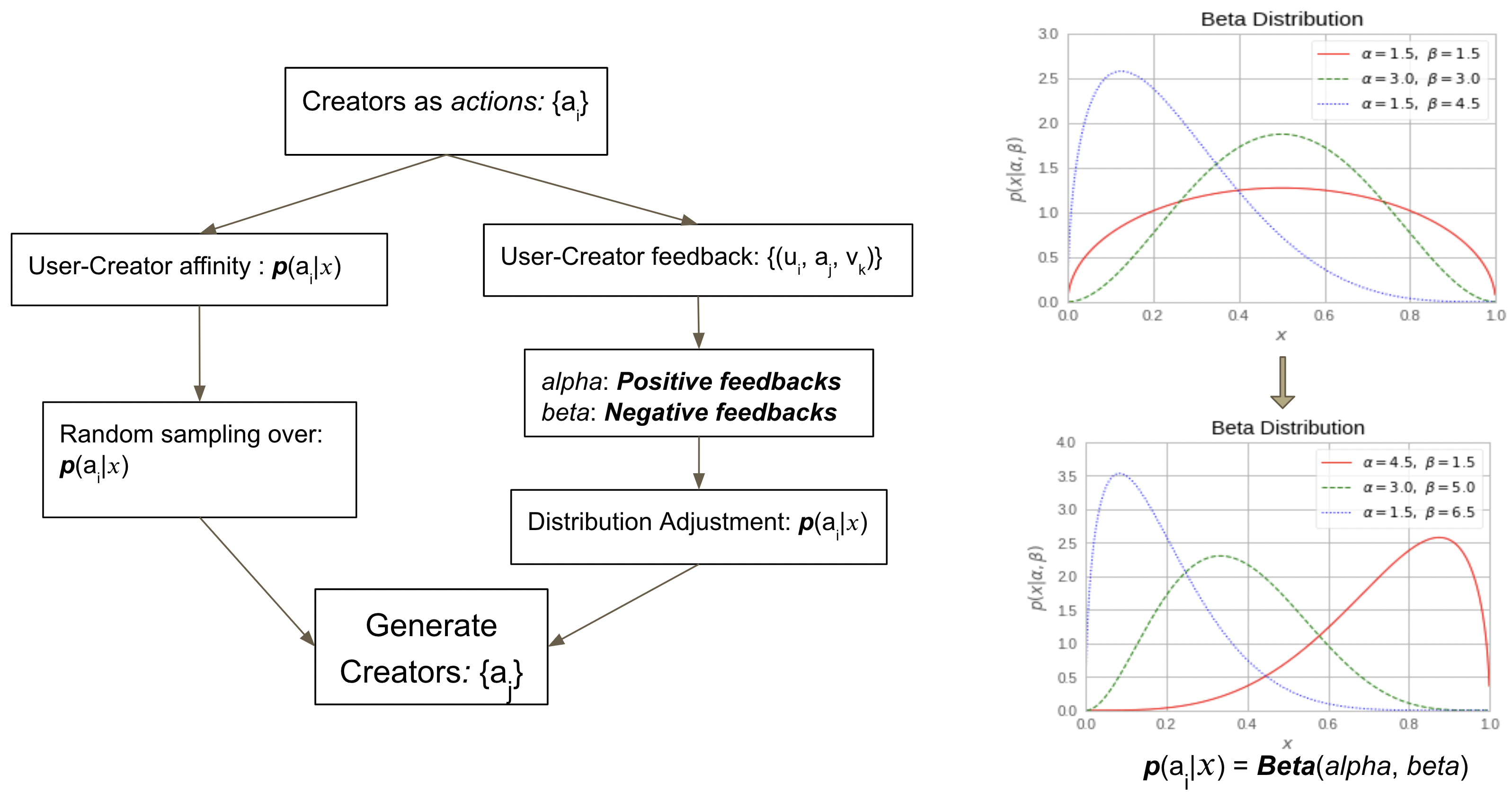}
  \caption{(Left) Random sampling vs Thompson sampling (Right) Distribution of sampling creators evolving over time}
  \Description{}
  \vspace{-3mm}
\end{figure}

\subsection{Delivery to balance explore/exploit} \label{Delivery-Balance}
In a point-wise ranking recommender system, introducing exploratory content is challenging as uncertain predictions usually get demoted by a well optimized ranker. We built a specialized delivery system to introduce uncertainty in a controlled manner in the recommendation system.

\textbf{Exploration Composition:} We used a probabilistic blending layer which considers composition of exploration in user's current session and slots an exploration video to enforce a predetermined composition

\section{Experiment}

\subsection{Setup}
To test our proposed approach, we perform an online experiment on Facebook Watch, a popular real-world video recommendation system serving billions of users. Inspired by \cite{chen2021google}, we split the users into four groups depending on two conditions.

Condition 1: whether to serve exploration content to a user. For users getting exploration content, we relying on our blending pass to control the impression from exploration videos to be roughly 6\%.

Condition 2: whether the recommender system is trained using exploration data. Two recommender models, model A and model B, are trained with same architecture and features. We remove the data collected through exploration from the training set of model A. To make a fair comparison, a certain amount of non-exploration data is randomly removed from model B’s training set so that model A and model B have the same size of training set.

Table~\ref{table:online_groups} above shows the condition combinations of the four groups.

\begin{table}
\captionsetup{font=small}
\begin{tabular}{c | c | c}
\toprule
Group & Exploration Content & Model  \\ 
\midrule
1       &   N   &  Model A                \\
2       &   Y   &  Model A                \\
3       &   N   &  Model B                \\
4       &   Y   &  Model B                \\
\bottomrule
\end{tabular}
\caption{Four groups with condition combinations used in online experiment.}
\label{table:online_groups}
\vspace{-10mm}
\end{table}

\subsection{Metrics}

When it comes to understanding the value of video exploration, one of the biggest challenges is finding the right metrics. Commonly used metrics like watch time can temporarily deteriorate as exploration leverages opportunity cost to show something with higher uncertainty. Moreover, we believe that exploration could take the recommender system out of local maxima and in to a global optimum, which is impossible without short-term regression on engagement metrics.

By considering that one of the main goals of exploration is to discover meaningful connections between users and their unknown content preferences, we define a set of metrics as proxies for such connections.

\begin{figure}[h]
  \centering
  \captionsetup{font=small}
 \includegraphics[scale=0.35]{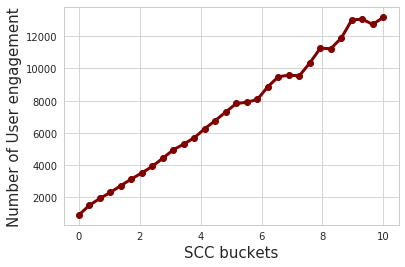}
  \caption{SCC showing strong correlation with future engagement}
  \Description{}\label{fig:scc}
\end{figure}

\textbf{Strong Creator Connection (SCC)}: SCC is generated when a user has at least $N$ engagement events with a particular creator in a window of $M$ days \footnote{\label{note1}We omitted the value for N, M, P here and below for business-compliance reasons.}. We use creators as our proxy for preferences/content, for creators generally publish content with a consistent genre, tone, theme, etc. In the short term, SCC is sensitive enough to reflect the connections built between users and creators. In the long run, we have seen a high correlation between SCC and user engagement (see Figure \ref{fig:scc}).

\textbf{Strong Creator Connection-Daily Active User (SCC DAU)}: This metric counts the number of users who has non-zero SCC metric. Increase of this metric indicates more users have built strong creator connection.

\textbf{Novel Strong Creator Connection (Novel SCC)}: Novel SCC is defined as strong creators connections formed between a user and a creator that this user has no engagement in last $P$ days\textsuperscript{1}. Comparing with SCC, Novel SCC further curves out the impact of exploration from non-exploration. It also helps to measure long term shift in user’s interest.

Aside from the SCC based metrics, we will also examine the commonly used engagement win metrics to measure the overall effectiveness of exploration framework.


\section{Results}

We conducted four weeks of A/B testing to rank videos for users on Facebook Watch using the propsed exploration framework. The test results show that exploration results in significant wins on both overall and novel strong creator connections.

\textbf{User Exploration Value}: Group 4 versus Group 1 shows the net impact of introducing exploration into the recommender system. It is a combined effect of users getting exploration content and the feedback of exploration information into the recommendation system. We observe an increase of 3.50\% in SCC, 0.26\% in SCC DAU and 0.85\% in Novel SCC. There is no statistically significant impact on users' video consumption or daily and weekly retention. Figure \ref{fig:SCC daily} below shows the SCC gains over the period.

\begin{figure}[h]
  \centering
  \captionsetup{font=small}
  \addtolength{\abovecaptionskip}{-5pt}
  \addtolength{\belowcaptionskip}{-5pt}
  \includegraphics[scale=0.14]{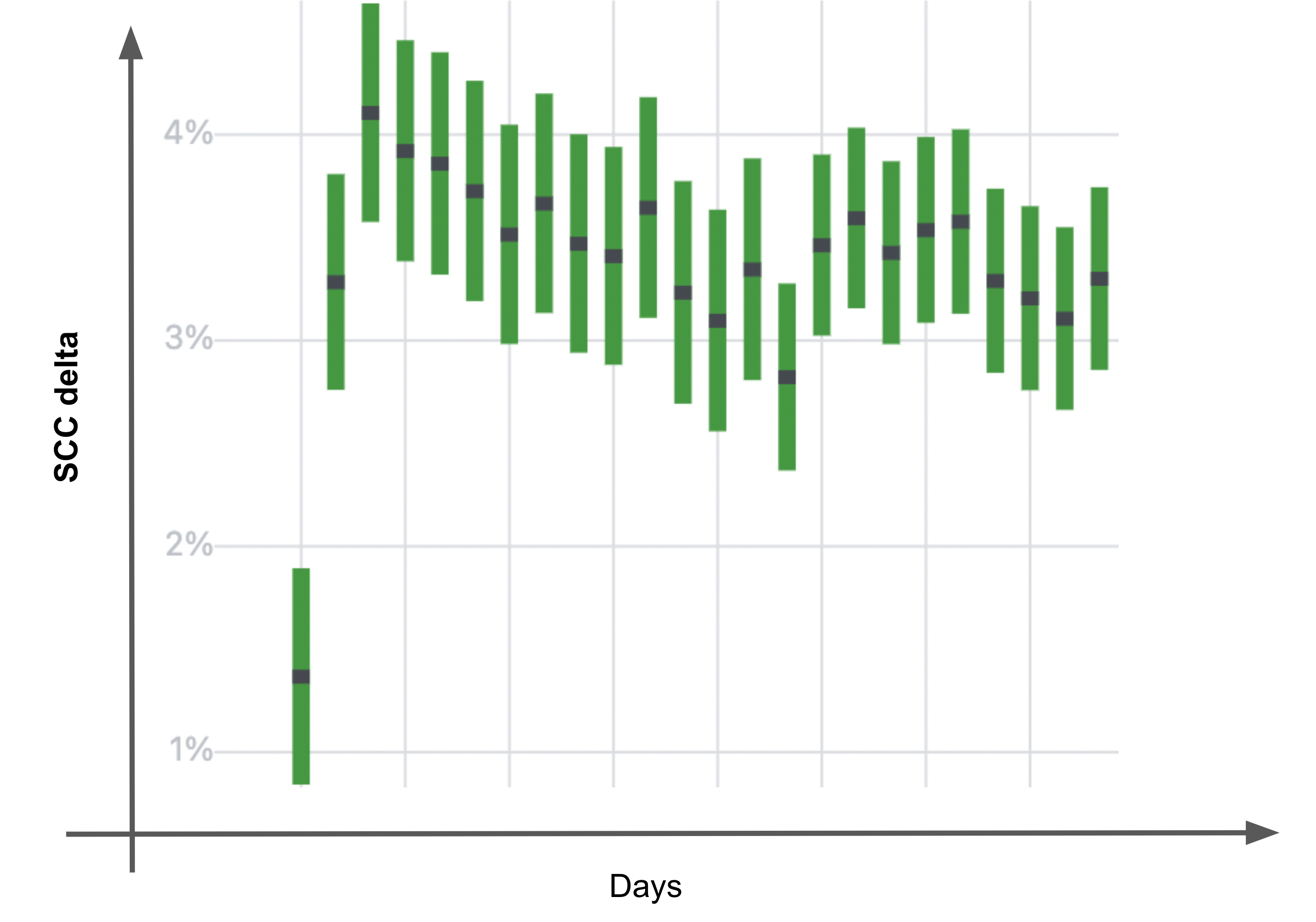}
  \caption{The SCC delta holds consistently in our A/B testing duration of four weeks}
  \Description{}
  \label{fig:SCC daily}
\end{figure}

\textbf{System Exploration Value}: Since neither Group 3 nor Group 1 serve exploration content, the comparison between Group 3 and Group 1 merely reflects the benefits of exploration data to the recommender system. We see a 0.28\% improvement in engagement with no statistically significant change in SCC metrics.

\textbf{Strict Exploration Value}: Group 2 vs Group 1 shows the impact of serving exploration content to users without adding a feedback loop to improve our recommendation model using information collected from exploration. It measures the net opportunity cost we incur to show more uncertain exploration content instead of videos from the more optimized existing recommendation system. We see an engagement regression of 0.53\%. On the other hand, serving exploration content generates a 0.55\% increase in SCC from the user-creator connection perspective even without the feedback loop.

\begin{table}
\centering
\captionsetup{font=small}
\begin{tabular}{|l|l|l|l|l|}
\hline
\begin{tabular}[c]{@{}l@{}}Model \\ Description\end{tabular}        & SCC    & \begin{tabular}[c]{@{}l@{}}SCC \\ DAU\end{tabular} & \begin{tabular}[c]{@{}l@{}}NOVEL \\ SCC\end{tabular} & Engagement \\ \hline
\begin{tabular}[c]{@{}l@{}}User \\ Exploration Value\end{tabular}   & \textbf{+3.50\%}  & \textbf{+0.26\%}                                              & \textbf{+0.85\%}                                                  & Neutral    \\ \hline
\begin{tabular}[c]{@{}l@{}}System \\ Exploration Value\end{tabular} & Neutral  & Neutral                                            & Neutral                                                 & \textbf{+0.28\%}     \\ \hline
\begin{tabular}[c]{@{}l@{}}Strict \\ Exploration Value\end{tabular}                                           & +0.55\% & Neutral                                             & +0.11\%                                                & \textbf{-0.53\%}    \\ \hline
\end{tabular}
\caption{Exploration metrics impact in A/B test relative to production.}
\label{table:online_results}
\vspace{-9mm}
\end{table}

\begin{figure}[h]
  \centering
  \captionsetup{font=small}
  \addtolength{\abovecaptionskip}{-5pt}
  \addtolength{\belowcaptionskip}{-7pt}
  \includegraphics[width=\linewidth]{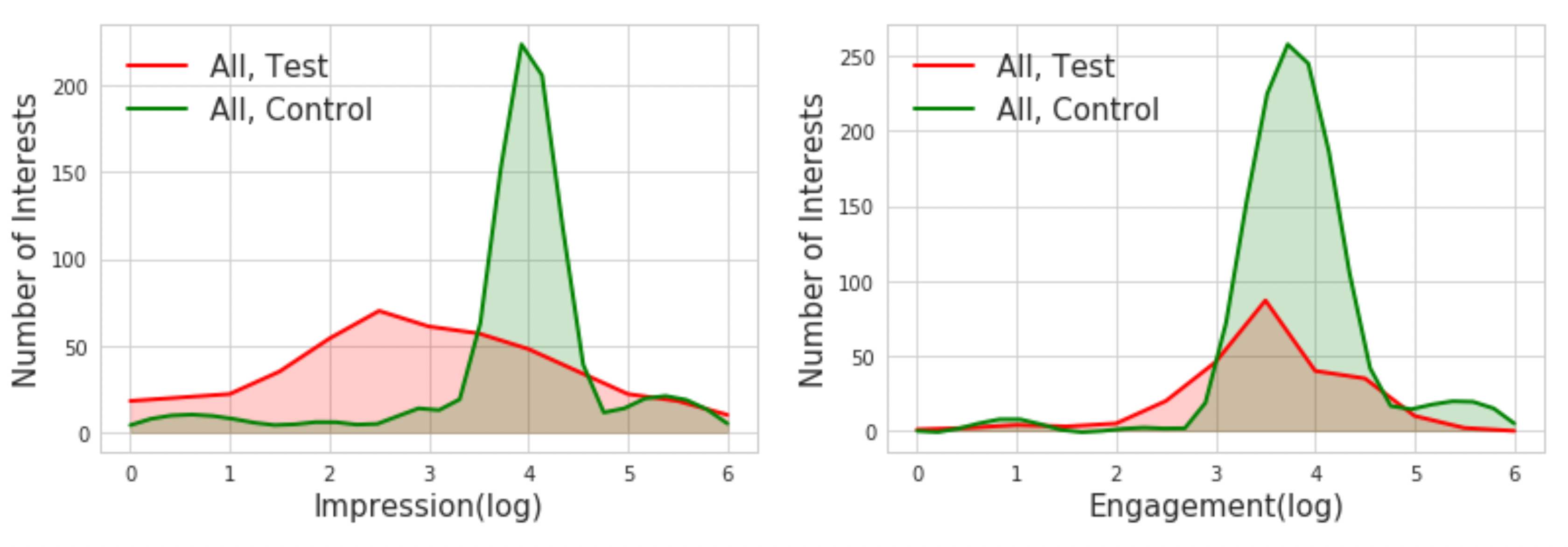}
  \caption{Interest distribution Impression (left) and Engagement (right)}
  \Description{}
  \label{fig:interest}
  \vspace{-3mm}
\end{figure}

\textbf{Distribution of Interests}: To visualize the impact of exploration on user engagement, we calculate user engagement for each subtopic separately and plot the histogram in Figure \ref{fig:interest}. The x-axis represents impressions and engagement in log scale, and the y-axis represents the number of subtopics in that bucket. Results with exploration (test) and without exploration (control) are plotted. The left figure shows that in the test group most subtopics have log impressions of around 4, while in the control group the impressions received by different subtopics varies much more. In other words, we are able to shift the distribution of interest impressions to a more balanced curve through exploration. Similarly, the shift in engagement distribution in the right figure indicates that relevant and novel interests can balance the engagement-interest distribution.

\section{Conclusion}

In this paper, we propose a framework for exploration on a large-scale recommender system serving billions of users. The framework can be easily integrated into an existing recommender system with minimal modifications and consists of three major steps:
\begin{enumerate*}[label=(\arabic*)]
\item Capturing user’s evolving interests expressed in terms of creators
\item Applying online exploration
\item A blending process that balances explore and exploit to ensure optimal prevalence of exploratory videos.
\end{enumerate*}
To capture the value of exploration, we also define a set of proxy metrics to measure meaningful user-creator connections. We ran extensive A/B tests to understand overall engagement and SCC metrics gains on our video platform. We also present various analysis showing how our exploration method helped capture users’ new interests by forming connections with new creators. Our work has been deployed in production on Facebook Watch, a popular video discovery and sharing platform serving billions of users.

\section{Future work}

The work described above has a number of interesting future directions. For example, the Personalized PageRank approach could be replaced by other methods to expand the exploration space, such as collaborative filtering or popularity-based methods. Furthermore, for online exploration, more complex approaches like Contextual Bandit could be applied to select relevant creators for users instead of the presented Thompson sampling technique. Alternatively, the slot-based mechanism for showing exploration videos at relevant positions could be migrated to a more dynamic system where exploration videos are shown at each position with a probability that depends on the user's preferences.

\bibliographystyle{ACM-Reference-Format}
\bibliography{main}

\end{document}